\documentclass{jfm}

\usepackage{amsmath}
\usepackage{color}
\usepackage{amssymb}
\usepackage{graphicx}
\usepackage{upgreek}

\newcommand{\tcr}[1]{\textcolor{black}{#1}}

\newcommand{\pd}[2]{\frac{\partial #1}{\partial #2}}

\newcommand{\ie}{\textit{i.e.}}

\graphicspath{ {./figures/} }

\usepackage{dcolumn}% Align table columns on decimal point
\usepackage{bm}% bold math
\usepackage{amsmath}
\usepackage{amssymb}
\usepackage{scalerel}[2014/03/10]
\usepackage{stackengine}

\usepackage[euler]{textgreek}

\usepackage{multirow}

\title[ ]{Impact of vaporization on drop aerobreakup}

\author{
 B. Boyd \aff{1}\corresp{\email{bradley.boyd@canterbury.ac.nz}}, S. Becker \aff{1}, \and Y. Ling\aff{2}\corresp{\email{stanley\_ling@sc.edu}}
}

\affiliation{
\aff{1}Department of Mechanical Engineering, University of Canterbury, Christchurch, New Zealand
\aff{2}Department of Mechanical Engineering, University of South Carolina, Columbia, South Carolina, USA
}

\begin{document}

\maketitle

		\begin{abstract} 
Aerodynamic breakup of vaporizing drops is commonly seen in many spray applications. While it is well known that vaporization can modulate interfacial instabilities, the impact of vaporization on drop aerobreakup is poorly understood. Detailed interface-resolved simulations were performed to systematically study the effect of vaporization, characterized by the Stefan number, on the drop breakup and acceleration for different Weber numbers and density ratios. It is observed that the resulting asymmetric vaporization rates and strengths of Stefan flow on the windward and leeward sides of the drop hinder bag development and prevent drop breakup. The critical Weber number thus generally increases with the Stefan number. The modulation of the boundary layer also contributes to a significant increase of drag coefficient. Numerical experiments were performed to affirm that the drop volume reduction plays a negligible role and the Stefan flow is the dominant reason for the breakup suppression and drag enhancement observed. 
		\end{abstract}
		
		%Use showkeys class option if keyword

%===============================================================================
%    Introduction
%===============================================================================
\section{Introduction}
Aerodynamic breakup of drops is a classical multiphase flow problem commonly encountered in various spray applications such as liquid fuel injection and spray cooling. When the relative velocity between the drop and the surrounding gas is sufficiently high, the drop deforms and may even break as it accelerates. In many atomization problems, bulk liquids first break into larger drops, which then undergo secondary atomization into even smaller child droplets. 
In the present study, the term \emph{aerobreakup} refers to any scenario in which the drop undergoes a topology change, such as forming a hole. For cases near the critical breakup condition, the drop may not fully atomize into a large number of smaller droplets. For theoretical and numerical studies, drop aerobreakup is often formulated in an idealized configuration, i.e., a stationary drop suddenly exposed to a uniform gaseous stream \citep{orourke_tab_1987, jain_secondary_2015, jain_secondary_2019, marcotte_density_2019, rimbert_spheroidal_2020}. Experimentally, this sudden change in relative velocity is achieved using a shock wave \citep{hsiang_drop_1995, theofanous_aerobreakup_2004} or a jet \citep{flock_experimental_2012, opfer_droplet-air_2014, jackiw_aerodynamic_2021}. The shape evolution and acceleration of the drop are governed by key dimensionless parameters, including the Weber ($\text{We}$), Ohnesorge ($\text{Oh}$), and Reynolds ($\text{Re}$) numbers, and the liquid-to-gas density ratio ($\eta$). Extensive experimental, theoretical, and numerical studies have been dedicated to investigating drop aerobreakup \citep{guildenbecher_secondary_2009, theofanous_aerobreakup_2011}. In recent years, high-fidelity experimental diagnostics and large-scale numerical simulations have emerged \citep{jackiw_aerodynamic_2021, chirco_manifold_2022, tang_bag_2023, ling_detailed_2023}, providing important details to understand the underlying physics of interfacial multiphase flow. 
While $\text{We}$ is often used to characterize different breakup modes, the effects of $\eta$ \citep{marcotte_density_2019, jain_secondary_2019}, $\text{Re}$ \citep{aalburg_deformation_2003, strotos_numerical_2016}, and heating \citep{strotos_aerodynamic_2016} may also be important for low-$\text{We}$ scenarios.

In applications where drop aerobreakup occurs in high temperature environments, such as injection of liquid fuel into combustion chambers, drop vaporization occurs simultaneously as the drop accelerates and deforms \citep{strotos_aerodynamic_2016}. Phase change is known to significantly influence canonical interfacial instabilities, such as Rayleigh-Taylor and Kelvin-Helmholtz instabilities \citep{hsieh_interfacial_1978, pillai_rayleightaylor_2018}. Since interfacial instability plays a significant role in drop deformation and breakup, it is expected that vaporization can also have a substantial influence on drop aerobreakup. However, a detailed characterization of the vaporization effect is lacking in the literature. The vaporization of drop liquid on the surface can be driven by heat transfer as a result of the temperature gradient or by mass transfer due to the vapor concentration gradient. This study focuses on the former scenario, where the rate of vaporization is determined by the imbalance of heat fluxes across the interface. This introduces additional parameters, such as the Prandtl ($\text{Pr}$) and Stefan numbers ($\text{St}$). Experimentally, the vaporization of a spherical drop in high-temperature environments has been studied, resulting in commonly used empirical relations for the drop vaporization rate \citep{renksizbulut_experimental_1983,abramzon_droplet_1989}. These correlations are strictly valid in the zero-$\text{We}$ limit and will significantly underestimate the vaporization rate for deformable drops with finite $\text{We}$ \citep{setiya_evaporation_2022, boyd_simulation_2023}. For drops with finite $\text{We}$ and $\text{St}$, the strong coupling between drop shape deformation and vaporization complicates the problem. On the one hand, significant drop deformation will change the gas flow around the drop and the temperature distribution in the thermal boundary layer near the surface, which leads to the time-varying and non-uniform distribution of the local vaporization rate on the drop surface. However, vaporization, in particular, the asymmetric vaporization rate on the windward and leeward sides of the drop, modulates the interfacial dynamics and drop acceleration, which in turn change the drop deformation and the breakup criteria. The goal of this study is to comprehensively investigate these complex interactions between drop aerobreakup and vaporization through detailed parametric numerical simulations. Beyond the commonly considered $\text{We}$, we will systematically vary two other key dimensionless parameters, \ie, $\text{St}$ and $\eta$, to fully characterize the vaporization effect.

%===============================================================================
%    Governing Equations
%===============================================================================
\section{Methodology}
To accurately predict the aerobreakup of vaporizing drops, two-phase interfacial flows with phase change must be faithfully resolved with the proper physical models and numerical methods. The simulation approach used for the present simulations has been presented in recent work of the authors \citep{boyd_consistent_2023,boyd_simulation_2023}, therefore, only a brief summary will be given here. The Navier-Stokes equations with surface tension are solved for two-phase flows using the one-fluid approach. 
The momentum and continuity equations are given as 
\begin{align}
	\rho \left( \frac{\partial \mathbf{u}}{\partial t}+ \mathbf{u} \cdot \nabla \mathbf{u} \right) & = - \nabla p + \nabla \cdot (2\mu \mathbf{D}) + \rho \mathbf{g}+ \sigma \kappa \delta_{\gamma} \mathbf{n}_{\gamma}\,,
	\label{eq:momentum}\\
	\nabla \cdot \mathbf{u} & = s_{\gamma} \left( \frac{1}{\rho_g}-\frac{1}{\rho_l} \right)\,,
    \label{eq:divergence}
\end{align}
where $\mathbf{u}$, $p$, $\mu$, $\rho$, $\sigma$, and $\kappa$ are the velocity, pressure, dynamic viscosity, density, surface tension coefficient, and interfacial curvature. The deformation tensor is defined as $\mathbf{D} = (\nabla \mathbf{u} + \nabla \mathbf{u}^T)/2$. The interface normal and the interface Dirac distribution function are denoted by $\mathbf{n}_{\gamma}$ and $\delta_{\gamma}$, respectively. The subscripts $l$ and $g$ refer to the liquid and gas properties, respectively, while the subscript $\gamma$ denotes properties related to the surface. 

The two different phases are distinguished by the liquid volume fraction, $f$, which follows the advection equation with the phase-change-induced source term, 
\ie, 
\begin{align}
	\frac{\partial f}{\partial t}+ \nabla \cdot \left( f \mathbf{u}\right ) = \frac{-s_{\gamma}}{\rho_l}\,.
	\label{eq:vof_advection}
\end{align}
In Eqs.~\eqref{eq:divergence} and \eqref{eq:vof_advection}, $s_{\gamma}$ is the mass flow rate per unit volume, which in turn depends on the mass flux at the interface ($j_{\gamma}$) and the interfacial area density ($\phi_{\gamma}$) as
\begin{align}
	s_{\gamma} = j_{\gamma} \phi_{\gamma}, 
	\label{eq:sm}
\end{align}
where $\phi_{\gamma}=A_{\gamma}/V_c$, where $A_{\gamma}$ is the liquid-gas interface area in a cell with a volume $V_c$.
The source term on the right of Eqs.~\eqref{eq:divergence} is responsible for the generation of the Stefan flow; while the counterpart on the right of \eqref{eq:vof_advection} accounts for the additional change in the interface location due to phase change. 

The phase change at the interface is driven by heat transfer, so $j_{\gamma}$ is determined by the gas-temperature gradient at the interface,
\begin{align}
	j_{\gamma} =\frac{1}{h_{l,g}}\left(k_l \nabla T|_{l,\gamma} \cdot \mathbf{n}_{\gamma} - k_g \nabla T|_{g,\gamma} \cdot \mathbf{n}_{\gamma}\right)\,,
	\label{eq:j_gamma}
\end{align}
where $T$, $k$, and $h_{l,g}$ are the temperature, thermal conductivity, and specific latent heat of vaporization, respectively. The gas and liquid temperature fields required to calculate $j_\gamma$ are obtained by solving the energy conservation equation for both the liquid and gas phases
\begin{align}
	& \rho_g C_{p,g} \left( \pd{T_g}{t} +  \mathbf{u} \cdot \nabla T_g  \right) = \nabla \cdot (k_g \nabla T_g) \, ,
	\label{eq:temp_gas}\,\\ 
	& \rho_l C_{p,l}\left(\pd{T_l}{t} +   \mathbf{u} \cdot \nabla T_l \right) = \nabla \cdot (k_l \nabla T_l) \, 
	\label{eq:temp_liq}
\end{align}
where $C_p$ is the isobaric-specific heat. Since $T_g$ and $T_l$ are only solved in the gas and liquid regions, the gas-liquid interface is treated as an embedded boundary where the temperature remains as the saturation temperature, $T_{\text{sat}}$.

To rigorously resolve the sharp and vaporizing interface, a novel volume-of-fluid (VOF) method was used \citep{boyd_consistent_2023, boyd_simulation_2023}. As the projection method is used to incorporate the continuity equation, the pressure Poisson equation is solved. The volumetric source due to vaporization was added to the pressure equation to account for the non-zero divergence of the velocity near the interface and the resulting Stefan flow. To avoid contaminating the velocity at the interfacial cells, the vaporization-induced volumetric source is distributed to the pure gas and liquid cells adjacent to the interface in a conservative manner. This treatment will guarantee that the velocities at the interface cells are correctly represented and can be used directly in VOF advection. The energy equations for both phases are solved with the Dirichlet boundary condition at the interface. To avoid calculating the gradient across the interface, we estimate the temperature gradient for each phase by extrapolating the neighboring pure cells for the corresponding phase. The contribution of vaporization to the motion of the interface toward the liquid side is accounted for by geometrically shifting the planar VOF interface. 

The physical model and numerical methods were implemented in the open source \emph{Basilisk} solver, and the solver has been thoroughly validated to accurately simulate two-phase interfacial flows, including the aero-breakup of drops without and with phase change \citep{zhang_modeling_2020, ling_detailed_2023, boyd_consistent_2023, boyd_simulation_2023}. A key feature of the \emph{Basilisk} solver is that the quadtree/octree mesh is used to discretize the domains, providing important flexibility to dynamically refine the mesh in user-defined regions. The adaptation criterion for the present study is based on the wavelet estimate of the discretization errors of the volume fraction, temperature, and velocity.
%===============================================================================
%  Results
%===============================================================================
\section{Results}
\label{section:results}
\subsection{Problem setup}
We consider a freely moving drop with diameter $D_0$, initially stationary and at saturation temperature in an unbounded domain, suddenly exposed to a uniform superheated stream of vapor of the drop liquid with temperature $T_\infty$ and velocity $U_\infty$, see Fig.~\ref{fig:3D_shapes}(a). 
Using the free-stream velocity  $U_\infty$, the initial drop radius $R_0=D_0/2$, where $D_0$ is the initial drop diameter, the liquid density $\rho_l$, and the difference between the free-stream and saturation temperatures ($T_\infty-T_{sat}$) as the characteristic scales, the dimensionless variables can be defined as, 
\begin{align}
	u^*=u/U_\infty\,, \quad 	x^*=x/R_0\,, \quad \rho^*=\rho/\rho_l\,, \quad T^*=(T-T_{sat})/(T_\infty-T_{sat})\,.
\end{align}
Following previous works on aerobreakup, time is non-dimensionalized by the drop breakup time \citep{ranger_aerodynamic_1969}, 
\begin{align}
	t^*=t U_{\infty} /(D_0 \sqrt{(\rho_l/\rho_g )})\,.
\end{align}
The definitions and values of the key dimensionless parameters, including the Weber ($\text{We}$), Stefan ($\text{St}$), Ohnesorge ($\text{Oh}$), Reynolds ($\text{Re}$) numbers, and the liquid-to-gas density ($\eta$) and viscosity ($m$) ratios, are provided in Table \ref{tab:parameters}.
The fluid properties are similar to those of acetone: the saturation temperature is $T_{sat}=359$ K, the surface tension is $\sigma=0.0153$ (N/m), the latent heat is $h_{l,g}=4.88\times 10^{5}$ J/kg, and the gas density, viscosity, and thermal conductivity are $\rho_g=5.11$ (kg/m$^{3}$), $\mu_g=9.59\times 10^{-6}$ Pa-s, $k_g=0.0166$ W/m-K, respectively.
In the present study, $\text{Re}=1000$, $\text{Pr}=0.84$, and $m=19.29$ are kept constant, while $\eta$, $\text{We}$ and $\text{St}$ are varied for parametric studies. $\eta$ is varied by changing $\rho_l$, and three values, \ie, $\eta=139$, 453 and 766, are considered, which correspond to the density ratios from acetone ($\eta=139$) to ammonia ($\eta=766$). By varying the free-stream temperature $T_{\infty}$, $\text{St}$ changes from 0 to 2. The range of $\text{We}$ considered is from 1 to 40, which covers the non-breakup and bag-breakup regimes and is sufficient to identify the critical Weber number, $\text{We}_{cr}$. To vary $\text{We}$ but to keep $\text{Re}$ fixed, $U_\infty$ and $D_0$ are varied simultaneously, as a result, $\text{Oh}$ will vary from $7\times10^{-4}$ to $0.01$, which is generally small and is not expected to influence the critical Weber number \citep{hsiang_drop_1995}. 

\begin{table}
\setlength{\tabcolsep}{3.5pt}
\centering
  \begin{tabular}{ccccccc} 
We			& St 	& Oh & Re & Pr	 &  $\eta$	& $m$\\ 
$\rho_g U_{\infty}^2 D_0/\sigma$ 	& $C_{p,g} (T_{\infty}-T_{sat})/ h_{l,g}$ &  ${\mu_l}/{\sqrt{\rho_l \sigma D_0 }}$ &  $\rho_g U_{\infty} D_0/\mu_g$ & $c_{p,g} \mu_g/k_g$ & $\rho_l/\rho_g$ 	& $\mu_l/\mu_g$\\
1 - 40	& 0 - 2	& 0.0007 - 0.01 	& 1000	& 0.84 & 139 - 766   & 19.29 \\ 
  \end{tabular}
  \caption{Key dimensionless parameters for the present simulations.}
  \label{tab:parameters}
\end{table}

Under the non-zero relative velocity between the drop and the vapor free stream, the drop is accelerated along the streamwise direction, while in the meantime it deforms and vaporizes. Both 2D-axisymmetric and 3D simulations were performed, in which adaptive quadtree/octree meshes were used. The resolution of the mesh is controlled by the level of refinement, $\mathcal{L}$, which corresponds to $2^\mathcal{L}$ cells in a coordinate direction. For 2D axisymmetric simulations, we have used $\mathcal{L}=13$ and the length of the computational domain is $L_0=32D_0$, which gives a minimum cell size equivalent to $\Delta=D_0/512$. For 3D simulations, we have used a coarser mesh, \ie, $\Delta=D_0/256$, due to the high computational cost. Grid refinement studies were conducted by varying $L=11$, 12, and 13 ($\Delta=D_0/128$, $D_0/256$, and $D_0/512$). The results, presented in Appendix~\ref{sec:grid_refinement}, confirm that the mesh resolution used is sufficient to accurately resolve the vaporization of a drop undergoing significant deformation.

\subsection{Vaporization-induced breakup suppression}
\begin{figure}
	\begin{center}
		\includegraphics [width=1.\columnwidth]{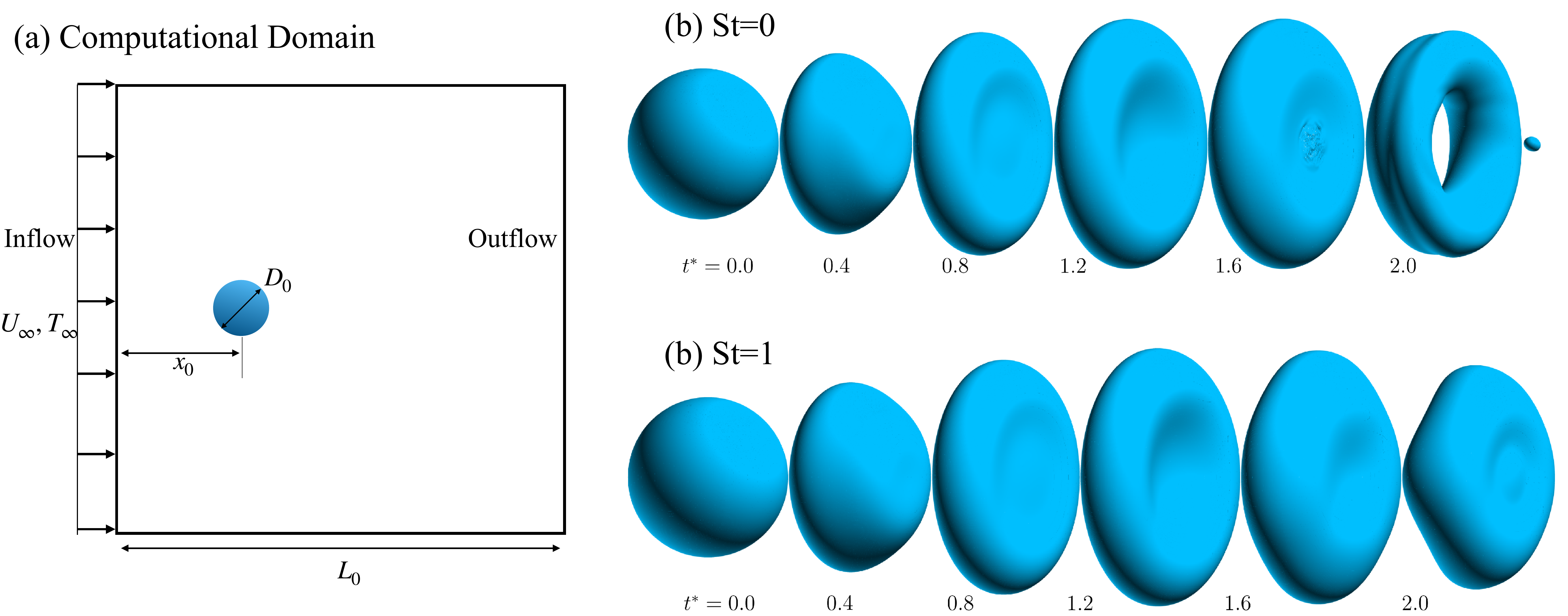}
	\end{center}
    \caption{(a) Computational domain for the 3D simulations and morphological evolutions for (b) non-vaporizing ($\text{St}=0$) and (c) vaporizing drops ($\text{St}=2$). The results are for $\eta=139$ and $\text{We}=10.5$.}
	\label{fig:3D_shapes} 
\end{figure}

The key observation from the simulation results is that drop vaporization tends to stabilize aerodynamic deformation and can suppress drop breakup. The strength of vaporization is characterized by $\text{St}$, and the vaporization rate increases with $\text{St}$. The 3D simulation results for non-vaporizing ($\text{St}=0$) and vaporizing ($\text{St}=2$) drops are shown in Fig.~\ref{fig:3D_shapes}(b) and (c), respectively, with all other parameters, including $\eta=139$ and $\text{We}=10.5$, kept unchanged. It can be observed that the non-vaporizing drop breaks at the center, while the vaporizing counterpart does not. The early-time shape evolutions for both drops are quite similar. The initially spherical drop expands radially into a disk shape ($t^*=0.8$) due to the pressure difference between the windward pole and the periphery of the drop \citep{rimbert_spheroidal_2020}. The internal radial flow within the liquid causes the axial thickness of the disk to reduce in time. The difference between the two cases arises around $t^* = 1.2$, when the periphery edge of the disk starts to retract toward the center, driven by surface tension. The radial retraction of the vaporizing drop is faster, and due to mass conservation, the thinning at the disk's center slows down, preventing the disk thickness from diminishing to zero. As a consequence, the two surfaces of the disk do not pinch to form a hole. In contrast, the non-vaporizing drop eventually undergoes a bag-mode breakup, characterized by the formation of a central hole, whereas the vaporizing drop maintains its initial topology without undergoing topological change.

We have performed the same cases using 2D axisymmetric simulations, which yield similar results for the present case, mainly due to the moderate $\text{Re}=1000$. The less expensive 2D simulations will then be performed for parametric simulations varying $\text{We}$, $\text{St}$, and $\eta$. 

\subsection{Modulation of drop surface}
\begin{figure}
	\begin{center}
		\includegraphics [width=1.\columnwidth]{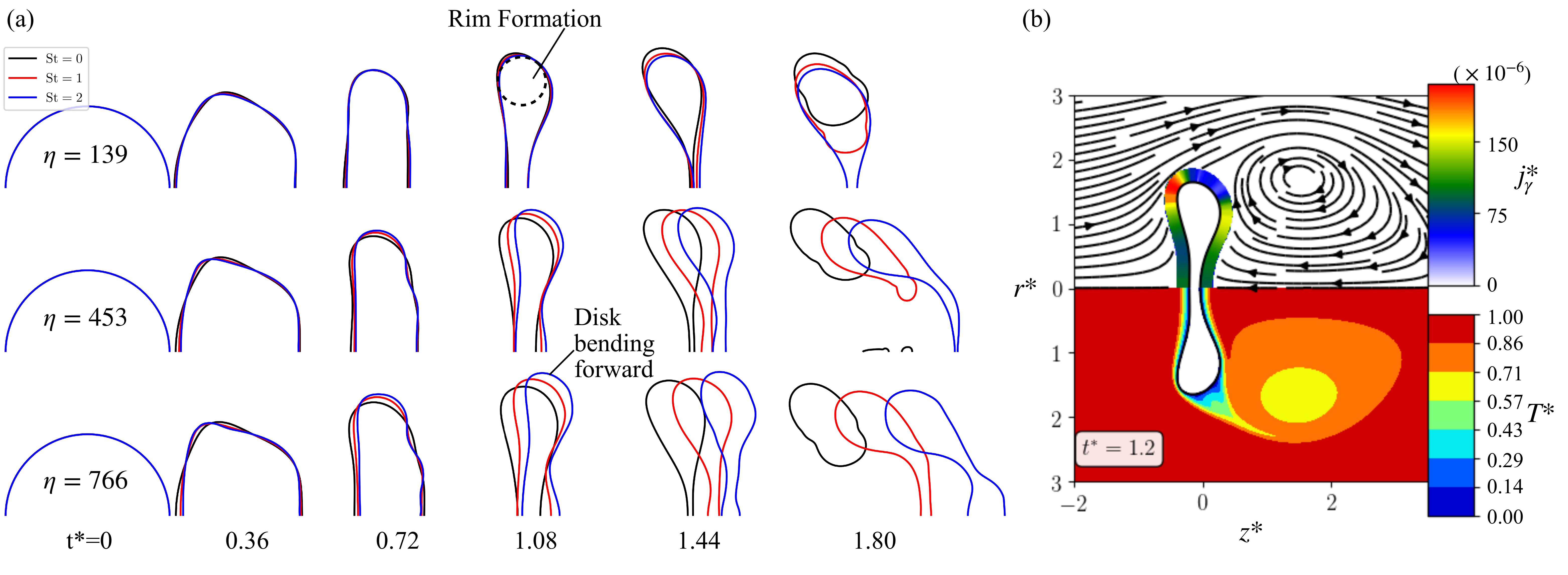}
	\end{center}
    \caption{(a) Temporal evolution of the drop surfaces for different $\eta$ and $\text{St}$. The results for $\eta=139$, 453, and 766 correspond to $\text{We}=10.5, 12$, and 12, respectively.  (b) The flow field and vaporization mass flux $j_\gamma^*=j_\gamma/(\rho_l U_\infty)$ (upper half) and temperature field ($T^*=(T-T_{sat})/(T_\infty-T_{sat})$) (lower half) at $t^*=1.2$ for $\eta=139$, $\text{We}=10.5$, and $\text{St}=1$.}
	\label{fig:compare_evolve} 
\end{figure}

To confirm that the vaporization-induced suppression of bag breakup is not a coincidence for the specific density ratio $\eta=139$ considered, additional cases $\eta=453$ and 766, are simulated. The time evolutions of the drop surfaces for different $\text{St}$ and $\eta$ are shown in Fig.~\ref{fig:compare_evolve}(a). When $\eta$ increases, the dynamics of drop deformation changes, and the critical Weber number that characterizes the onset of drop breakup, $\text{We}_{cr}$, decreases. The decrease in $\text{We}_{cr}$ with $\eta$ is significant when $\eta \gtrsim 100$ \citep{marcotte_density_2019}. For the three different $\eta$ considered here, we have used $\text{We}=12$ for $\eta=453$ and 766, and $\text{We}=10.5$ for $\eta=139$, which are slightlly over the corresponding $\text{We}_{cr}$. If there is no vaporization ($\text{St}=0$), drops of all $\eta$ will break in the bag mode. 

Figure \ref{fig:compare_evolve}(a) clearly shows that the vaporization-induced breakup suppression is present for all $\eta$. The difference between the interface shapes for different $\text{St}$ is actually larger when $\eta$ increases. As a result, the detailed mechanisms behind the stabilizing effect of vaporization are better revealed in the cases with higher $\eta=453$ and 766. After the drop deforms from a sphere to a disk, a rim is formed at the periphery, see $t^*=1.08$ in Fig.~\ref{fig:compare_evolve}(a). It is observed that the rim moves downstream faster, causing the disk to bend forward, and the level of bending increases with $\text{St}$ and $\eta$. This modulation of the drop shape is due to the asymmetric vaporization rates on the windward and leeward sides of the rim, see Fig.~\ref{fig:compare_evolve}(b). Since the local vaporization mass flux, $j_\gamma$, is non-zero only in the interfacial cells (cells with fractional values of VOF), we have graphically ``thickened" the non-zero $j_\gamma$ region in the top half of Fig.~\ref{fig:compare_evolve}(b) for better visualization, from which different vaporization rates on the two sides of the rim can be clearly seen. The vaporization rate depends on the temperature gradient near the interface, which is in turn inversely proportional to the thickness of the thermal boundary layer near the interface, given $T_{\infty}-T_{sat}$ is fixed. Due to the stagnation flow on the windward side, the boundary layer thickness is much smaller than that on the leeward side where the flow separates, see Fig.~\ref{fig:compare_evolve}(b). As a result, the vaporization is significantly stronger on the windward side of the rim. Due to the density difference between the liquid and the vapor, Stefan flow is induced at the interface, which repels the interface in the opposite normal direction. The higher vaporization rate on the windward side of the rim leads to a stronger repelling force, which causes the disk to bend forward. 

The excessive forward bending of the disk hinders bag formation and development. As shown in previous studies on bag breakup \citep{ling_detailed_2023, tang_bag_2023}, the bag forms when the center of the disk becomes sufficiently thin, as seen at $t^*=1.44$ for $\eta=766$ and $\text{St=1}$ in Fig.~\ref{fig:compare_evolve}(a). The center of the disk moves downstream faster than the periphery, forming to a bag with an upstream-facing opening, as observed at $t^*=1.80$ for $\eta=766$ and $\text{St}=1$. Vaporization, however, modulates the drop shape in the opposite manner, causing the edge to move downstream faster than the center, forming a ``reverse bag" with a downstream-facing opening, as seen at $t^*=1.08$ for $\eta=766$ and $\text{St}=2$. This modulation hinders bag development. Near $\text{We}_{cr}$, the force driving bag growth (the pressure difference across the drop) is only slightly higher than the resisting force (the surface tension on the periphery). This unfavorable modulation by vaporization is sufficient to suppress bag development and prevent hole formation in the bag.

As the Stefan flow is related to the density difference between vapor and liquid, Stefan flow and the resultant repelling force are enhanced when $\eta$ increases. For the highest $\eta=766$, vaporization results in a more significant modulation of the shape and drag of the drop. As a result, even a smaller $\text{St}=1$ is sufficient to suppress drop breakup, in contrast to the cases for $\eta=139$ where suppression is not observed until $\text{St}=2$.

\subsection{Modulation on gas flow and drag}
\begin{figure}
	\begin{center}
		\includegraphics [width=1.\columnwidth]{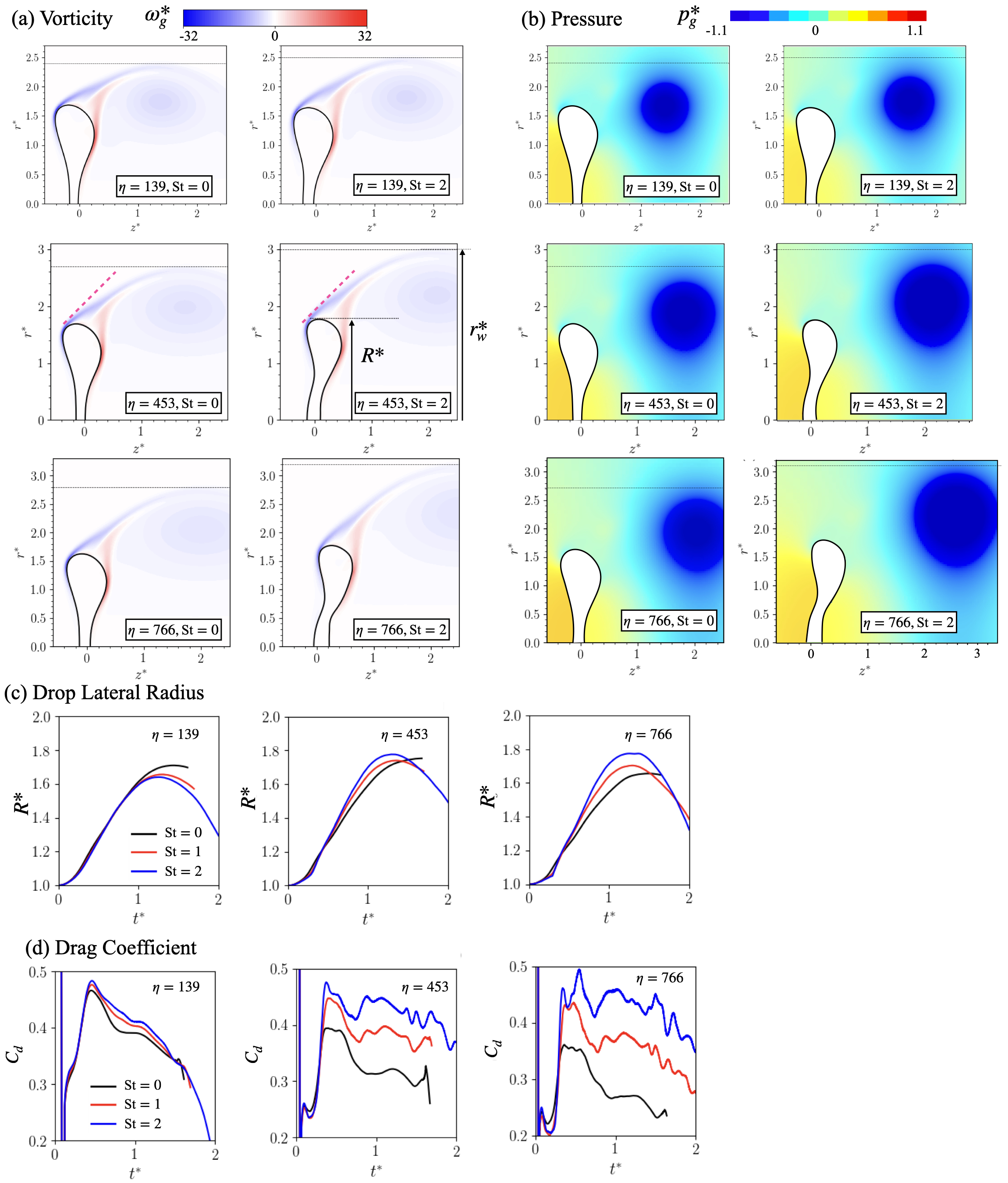}
	\end{center}
    \caption{The gas (a) vorticity ($\omega^*_g=\omega_g R_0/U_\infty$) and (b) pressure ($p^*_g=p_g /(\rho_g U_\infty^2)$) fields at $t^*=1.2$ for different density ratios $\eta=139$, 453, \tcr{766}, and Stefan numbers $\text{St=0}$ and 2. The temporal evolutions of the drop lateral radius $R^*$ and the drag coefficient $C_d$ for different $\eta$ and $\text{St}$ are shown in (c) and (d), respectively. The dashed lines in (a) and (b) indicate the lateral radius of the wake ($r_w^*$) estimated based on vorticity.} 
	\label{fig:wake_drag} 
\end{figure}

Another important effect of vaporization is the enhancement of drag and drop acceleration, which can be qualitatively observed in Fig.~\ref{fig:compare_evolve}(a). Detailed quantitative measurements of the drag coefficient ($C_d$) are given in Fig.~\ref{fig:wake_drag}(d). Here, the drag coefficient is defined based on the instantaneous drop frontal area,
\begin{equation}
C_d=\frac{F_d}{\frac{1}{2}\rho_g (U_\infty-u_d)^2 \pi R^2} = \frac{F_d^*}{\frac{1}{2\eta}(1-u_d^*)^2 \pi (R^*)^2},
\label{eq:Cd}
\end{equation}
where $F_d^*=F_d/(\rho_l U\infty^2 R_0^2)$. 
While $C_d$ varies over time, it is observed that $C_d$ generally increases with $\text{St}$; however, the detailed mechanisms behind this trend remain to be investigated.

Since vaporization on the two sides of the rim is asymmetric, the Stefan flow and the resulting repelling force are stronger on the upstream side of the rim, leading to a net force in the stream-wise direction, which contributes to an increase in drag in the rim region. However, this contribution to the overall drag is expected to be minor, as vaporization on the drop surfaces away from the edge rim is approximately symmetric, see Fig.~\ref{fig:compare_evolve}(b). Another possible reason for the higher drag coefficient is the larger lateral drop radius when vaporization is present, as seen in the results for $\eta=453$ in Figs.~\ref{fig:compare_evolve}(a) and (c). While the drag on the drop will increase with $R^*$ and the frontal area, the drag coefficient ($C_d$), see Eq.~\ref{eq:Cd}, will not. Interestingly, for $\eta=139$, $R^*$ actually decreases slightly with $\text{St}$, see $t^*=1$, while $C_d$ remains higher for larger $\text{St}$. Therefore, the increase in $C_d$ when $\text{St}$ increases, as observed in Fig.~\ref{fig:wake_drag}(d), cannot be explained by the modulation of $R^*$ alone.

The key mechanism for drag enhancement is the modulation of the wake structure by the Stefan flow. It can be clearly observed from the vorticity fields (Fig.~\ref{fig:wake_drag}(a)) that the maximum lateral radius of the wake, $r_w^*$, increases when vaporization is present. Near the flow separation point, the Stefan flow pushes the vorticity layer away from the interface, resulting in a larger slope of the wake boundary and eventually in a larger $r_w^*$. The pressure in the wake is generally low, as shown in Fig.~\ref{fig:wake_drag}(b), and thus the enlargement of the wake radius contributes to the increase of the low-pressure region area and the net drag acting on the drop. This mechanism is similar to the drag crisis for a solid sphere, in which the drag is reduced when the wake lateral radius decreases due to the delay of the turbulent boundary layer separation \citep{tiwari_flow_2020}. 
\tcr{Since the strength of the Stefan flow is enhanced as $\eta$ increases, its modulation of the wake becomes more significant for higher $\eta$, as shown in Figs.~\ref{fig:wake_drag}(a) and (b). As depicted in Fig.~\ref{fig:wake_drag}(a), when $\text{St}$ increases from 0 to 2, $r_w^*$ for $\eta=139$ increases from 2.4 to 2.5, while for $\eta=453$ and 766, $r_w^*$ increases from 2.7 to 3.0 and from 2.8 to 3.3, respectively. Correspondingly, the vaporization-induced enlargement of the low-pressure region for $\eta=766$ is more pronounced than for $\eta=139$, see Fig.~\ref{fig:wake_drag}(b). Consequently, the increase in $C_d$ with $\text{St}$ is more significant for higher $\eta$. When $\text{St}$ increases from 0 to 2, $C_d$ at $t^*=1$ for $\eta=766$ increases by approximately 67\%, whereas the increases for $\eta=453$ and 139 are about 40\% and 5\%, respectively, see Fig~\ref{fig:wake_drag}(d).}

\subsection{Effects of Stefan flow and interface receding}
The drop vaporization results in two key modulations in interfacial dynamics: (1) the interface receding towards the liquid, reducing the drop volume, and (2) the Stefan flow induced by the fluid volume generated near the interface that occurs when a high-density liquid turns into a low-density vapor. While the previous discussions are focused on the effect of the Stefan flow,  it remains to affirm that it indeed dominates the other contribution in the breakup suppression and drag enhancement. For this purpose, additional  numerical ``experiments" were performed to investigate the individual effect on these two features. 

\begin{figure}
	\begin{center}
		\includegraphics [width=1.\columnwidth]{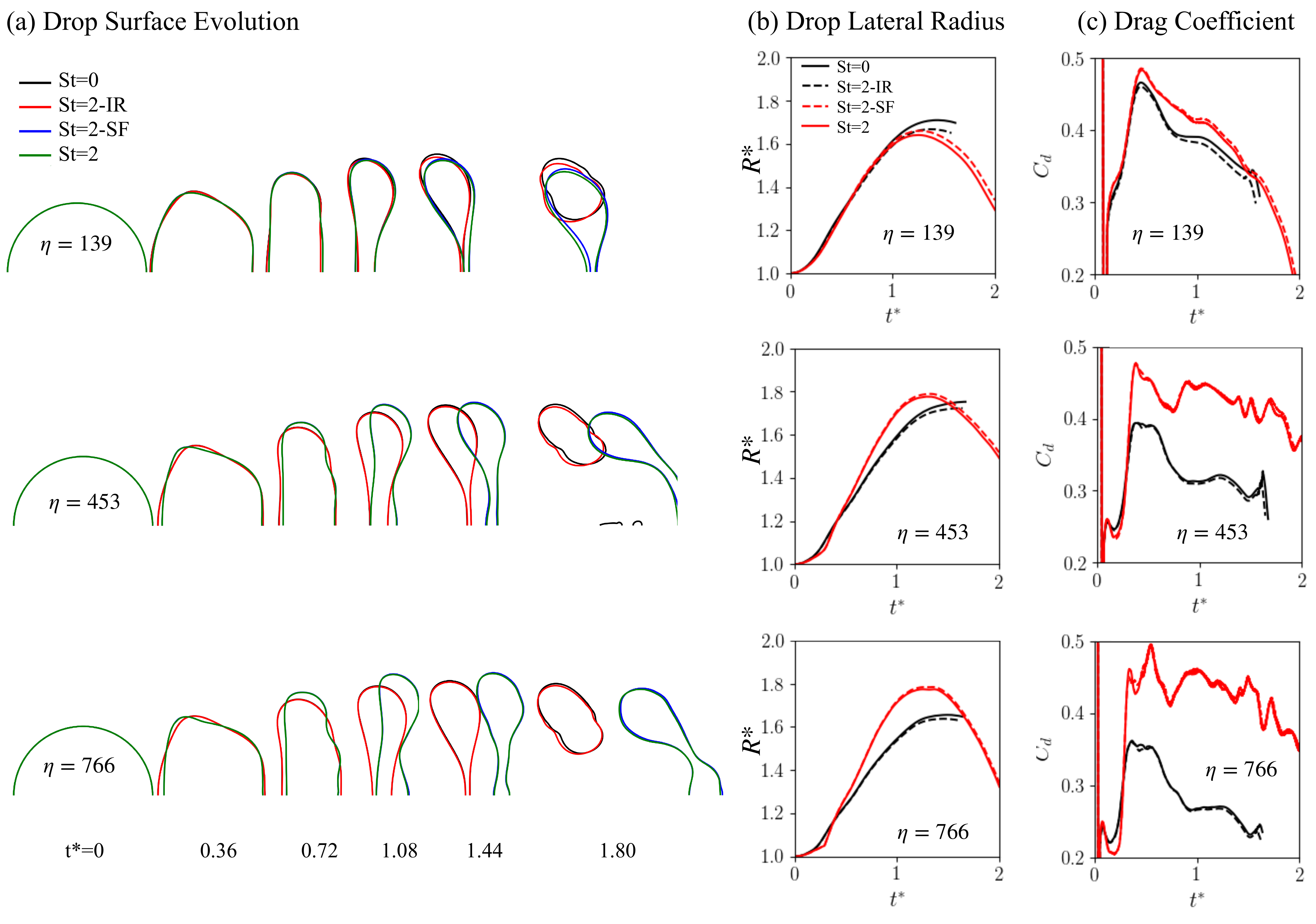}
	\end{center}
    \caption{Temporal evolutions of (a) drop surfaces, (b) drop lateral radius $R^*$ and (c) drag coefficient $C_d$ for different numerical tests, \ie, $\text{St}=0$ represents cases without vaporization; $\text{St=2}$-IR represents the numerical tests where the interface recedes due to vaporization but the source term on the right of Eq.~\eqref{eq:divergence} is turned off to avoid the production of the Stefan flow; $\text{St=2}$-SF represents the numerical tests where the Stefan flow is generated due to vaporization, but the source term on the right of Eq.~\eqref{eq:vof_advection} is set to zero to avoid interface receding toward the liquid phase due to vaporization; $\text{St}=2$ represents a full simulation with both interface-receding and Stefan flow effects accounted).}
	\label{fig:Stefan_flow_effect} 
\end{figure}

Four numerical tests are performed and the results are shown in Fig.~\ref{fig:Stefan_flow_effect}: (1) $\text{St=0}$, where there is no vaporization; (2) $\text{St=2}$-IR (IR: Interface Receding), where the interface recedes but Stefan flow is turned off; (3) $\text{St=2}$-SF (SF: Stefan Flow), where Stefan flow is produced but the interface receding is turned off, so the drop volume remains unchanged; and finally, (4) $\text{St=2}$, where the full vaporization effect is incorporated. In the simulation for $\text{St=2}$-IR, the source term on the right of Eq.~\eqref{eq:divergence} is set to zero, while the source term on the right of Eq.~\eqref{eq:vof_advection} remains unchanged. As a result, the VOF interface recedes toward the liquid side due to vaporization, but Stefan flow is not produced. For $\text{St=2}$-SF, the opposite is done: the source term on the right of Eq.~\eqref{eq:vof_advection} is set to zero while the counterpart for Eq.~\eqref{eq:divergence} remains unchanged. Consequently, Stefan flow is induced, but the interface does not recede due to vaporization, and the drop volume remains constant. It should be noted that these two cases are ``unphysical" and can only be examined in a numerical study by switching vaporization-related source terms on and off in different equations. However, these cases are very helpful for understanding the individual effects of vaporization and identifying the dominant one.

For both $\eta$, the drop surfaces for $\text{St=2}$-IR are very similar to those for $\text{St}=0$ throughout the drop evolution, see Fig.~\ref{fig:Stefan_flow_effect}(a). Holes are formed for both cases, leading to a change in the drop topology. This indicates that vaporization-induced recession of the interface and resulting volume change have minimal impact on drop deformation. In contrast, the drop surfaces for the case $\text{St=2}$-SF with Stefan flow turned on but with the interface receding turned off are close to the full simulation results for $\text{St}=2$ for all times, and eventually the breakup is suppressed for both cases. The results for the drop lateral radius shown in Fig.~\ref{fig:Stefan_flow_effect}(b) are consistent, the evolutions of $R^*$ for $\text{St}=0$ and $\text{St=2}$-IR are similar, while the results for $\text{St=2}$-SF agree well with those for $\text{St}=2$. Therefore, we can conclude that the Stefan flow is the primary effect of vaporization that influences the drop deformation. 

The drag coefficient for the numerical tests is presented in Fig.~\ref{fig:Stefan_flow_effect}(c). The good agreement between the results for $\text{St=2}$-SF for $\text{St}=2$ confirms that the Stefan flow is responsible for the drag enhancement. Without incorporating the Stefan flow, as in the test $\text{St=2}$-IR, one will significantly underestimate the drag and drop acceleration.

\subsection{Impact on breakup criteria}
The drop breakup criteria are often defined based on the critical Weber number, $\text{We}_{cr}$. When $\text{We}$ is lower than $\text{We}_{cr}$, the drop will remain unbroken throughout the acceleration. Parametric 2D axisymmetric simulations have been performed to identify $\text{We}_{cr}$ for different $\eta$ and $\text{St}$. We have considered 3 values of $\eta$ and 5 values of $\text{St}$, so there are 15 cases with different $\eta$-$\text{St}$ combinations. For each case, simulations of different $\text{We}$ are run to find $\text{We}_{cr}$ iteratively, using a bisection method with the minimum and maximum limits of $\text{We}=1$ and 40, respectively. The tolerance for iteration convergence is 0.2, that is, the root finding error for $\text{We}_{cr}$ is less than 0.2. For each case, about 5 to 7 simulations are needed to reach the desired tolerance. For all 15 cases, more than 100 runs were performed. 

For some cases just above $\text{We}_{cr}$, the drop lateral radius $R^*$ is decreasing when the drop breaks at the center. Those drops will eventually return to the ellipsoidal shape and will not atomize into a large number of child droplets. Such a near-$\text{We}_{cr}$ behavior has been identified in previous studies as well \citep{ling_detailed_2023}. We considered the formation of a hole in the drop bag to also be a breakup case to give a robust criterion for determining $\text{We}_{cr}$. 

\begin{figure}
	\begin{center}
		\includegraphics [width=0.75\columnwidth]{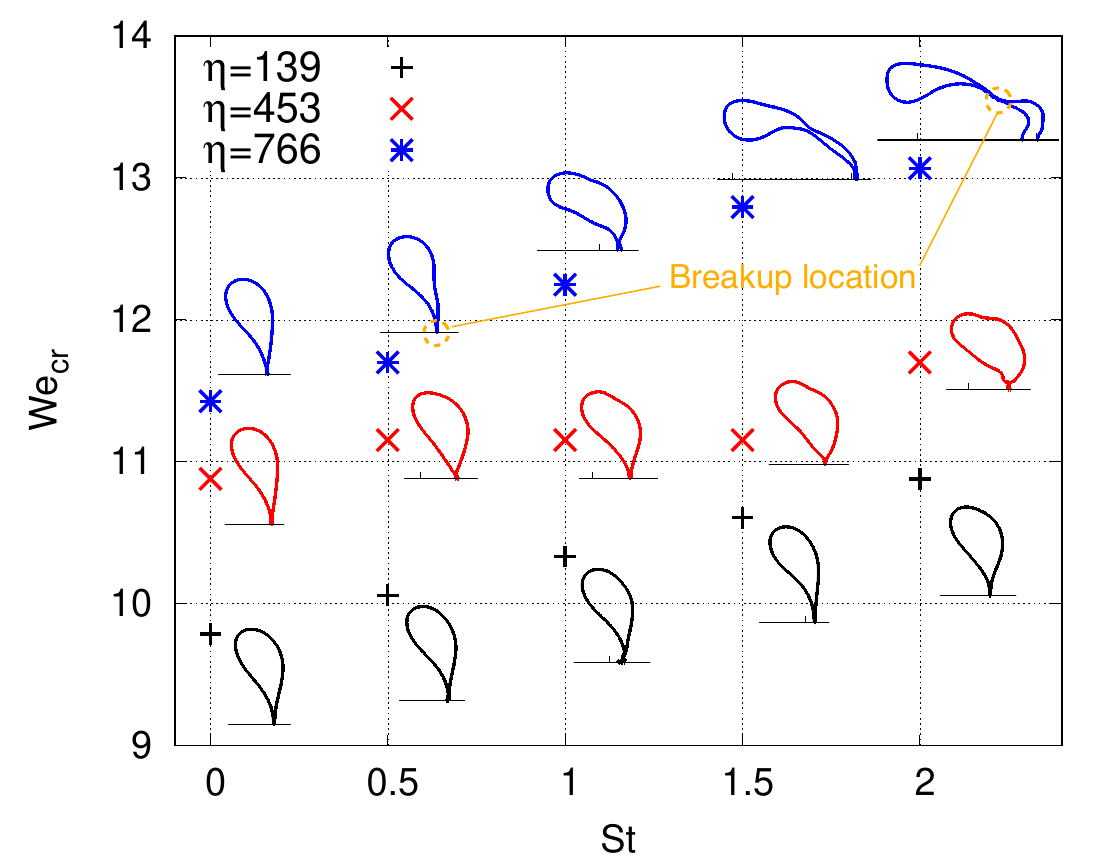}
	\end{center}
\caption{The critical Weber number $\text{We}_{cr}$ for the aerodynamic breakup of a vaporizing drop as a function of the Stefan number $\text{St}$, for different liquid-to-vapor density ratios ($\eta=139$, 453, and 766). The corresponding drop surfaces upon breakup are also shown.}
	\label{fig:we_critical} 
\end{figure}

The results of $\text{We}_{cr}$ for different $\eta$ and $\text{St}$ are summarized in Fig.~\ref{fig:we_critical}. \tcr{
For non-vaporizing drops ($\text{St}=0$), $\text{We}_{cr}=9.8$, 10.9, and 11.4 for liquid-to-gas density ratios $\eta=139, 453$, and 766, respectively. Experimental measurements for $\mathrm{We}_{cr}$ available in the literature are mainly for non-vaporizing drops with large $\eta$, and $\mathrm{We}_{cr}\approx 11$ for water drops in air ($\eta=831$) \citep{guildenbecher_secondary_2009}. Therefore, our results for $\mathrm{We}_{cr}$ agree well with the experimental data in the literature. Furthermore, the present results show that $\text{We}_{cr}$ decreases slightly with $\eta$, which is also consistent with previous studies on the density ratio effect \citep{marcotte_density_2019, jain_secondary_2019}. To the best knowledge of the authors, there are no experimental measurements for $\mathrm{We}_{cr}$ for vaporizing drops.
}
Since vaporization tends to stabilize the drop, $\text{We}_{cr}$ generally increases with $\text{St}$, meaning a higher gas dynamic pressure is required to break the drop. It was found that $\text{We}_{cr}$ increases by approximately 13\% for $\eta=766$ when $\text{St}$ increases from 0 to 2.

The drop shapes upon breakup at the corresponding $\text{We}_{cr}$ are also shown in Fig.~\ref{fig:we_critical}, where it can be observed that vaporization modifies the drop shape at the point of breakup. As expected, the breakup outcomes, including the velocity and size distributions of the resulting child droplets, will also change. However, a detailed study of the child droplet statistics is beyond the scope of the present work.

Although the general increasing trend of $\text{We}_{cr}$ with $\text{St}$ holds for all density ratios, a closer examination reveals that the variations in $\text{We}_{cr}$ depend on $\eta$. For $\eta=139$, $\text{We}_{cr}$ increases approximately linearly as $\text{St}$ increases, and the drop shapes remain similar. In contrast, the variation of $\text{We}_{cr}$ for higher $\eta$ is more complex. Specifically, for $\eta=453$, $\text{We}_{cr}$ shows little change within the searching tolerance of 0.2 for $\text{St}$ values between 0.5 and 1.5. In these cases, vaporization delays the breakup and modifies the drop shape at the point of breakup, but it is not sufficient to suppress the breakup. For $\eta=453$, the breakup time for $\text{St}=1.0$ is $t^*=1.907$, which increases to 1.925 for $\text{St}=1.5$, while $\text{We}_{cr}$ for these two cases remains almost the same.

For $\eta=766$, although $\text{We}_{cr}$ increases monotonically with $\text{St}$, similar to the trend observed for $\eta=139$, the drop shape upon breakup varies more significantly with changes in $\text{St}$. While the shape for the non-vaporizing case ($\text{St}=0$) at $\eta=766$ closely resembles that of its lower-$\eta$ counterparts, the drop experiencing strong vaporization, such as at $\text{St}=2$, displays a very different breakup shape, deforming into a long bag along the streamwise direction. For low $\text{St}$ values, the drop at $\eta=766$ breaks in a typical bag mode by pinching the two interfaces and forming a hole near the axis, similar to what occurs for lower $\eta$ values, which is expected when $\text{We} \approx \text{We}_{cr}$. However, as $\text{St}$ increases, the pinching location shifts away from the axis, and the breakup morphology appears to shift toward a bag-stem mode, though a more detailed investigation would be required to confirm this observation.

Finally, it is noted that the variation of $\text{We}_{cr}$ discussed above are strictly valid for the current range of parameters considered, namely $139\le \eta \le 766$ and $0\le \text{St} \le 2$. Extension of the present study to higher $\text{St}$ and ambient temperature will be relegated to our future work. 

%===============================================================================
%  Conclusions
%===============================================================================
\section{Conclusions}
\label{section:conclusions}
Parametric 3D and 2D axisymmetric interface-resolved simulations are conducted in the present study to investigate the aerobreakup of vaporizing drops. The Weber and Stefan numbers, and the liquid-to-gas density ratio, are varied to systematically study the effect of vaporization on the drop deformation/breakup and drag. For a wide range of liquid-to-density ratios, vaporization contributes to hindering the drop deformation and, if the Stefan number is sufficiently high, can suppress bag breakup. The vaporization also contributes to enhancement of the drag: it is shown that the drag coefficient for a vaporizing drop can be 84\% higher than its non-vaporizing counterpart. The Stefan flow induced by the interface vaporization plays the dominant role in modulations of the drop deformation stabilization and drag enhancement. The asymmetric vaporization rates on the windward and leeward sides of the edge rim and the resulting different strength of Stefan flow cause the disk to deform in a opposite manner compared to the bag, which hinders the bag development and breakup. The Stefan flow modifies the boundary layer dynamics near the separation point, leading to a wider wake and a large drag. More than one hundred simulations were run to identify the critical Weber number for different Stefan numbers and density ratio. When the Stefan number increases from 0 to 2, the critical Weber number can increase \tcr{up to} 13\%.

%===============================================================================
%  Appendix
%===============================================================================
\appendix
\section{Grid refinement studies}
\label{sec:grid_refinement}
\begin{figure}
	\begin{center}
		\includegraphics [width=0.8\columnwidth]{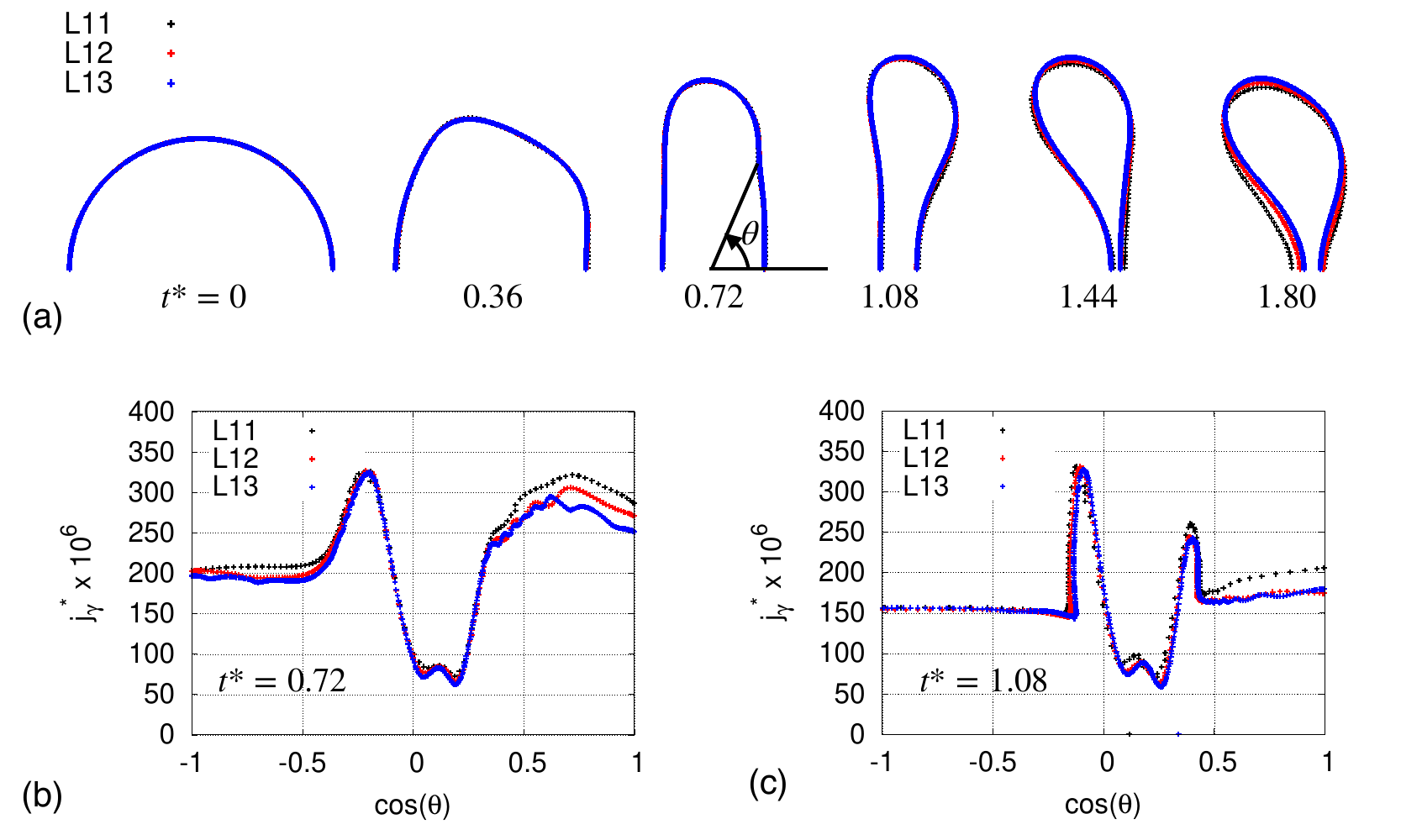}
	\end{center}
    \caption{(a) Temporal evolution of the drop surfaces for different mesh refinement levels $L=11$, 12, and 13, for the case $\eta=139$, $\text{We}=10.5$ and $\text{St}=2$. (b) and (c) Distribution of the vaporization mass flux $j_\gamma$ on the drop surface for $t^*=0.72$ and 1.08, respectively. }
	\label{fig:grid_refinement} 
\end{figure}

Grid refinement studies for non-vaporizing drops can be found in our previous paper \citep{ling_detailed_2023}. The results of grid refinement studies for a vaporizing drop ($\eta=139$, $\text{We}=10.5$, and $\text{St}=2$) are shown in figure \ref{fig:grid_refinement}. This case represents a drop with strong vaporization and high $\text{St}$ in the present study. Three different refinement levels, $L=11$, 12, and 13, corresponding to 128, 256, and 512 cells per initial drop diameter, are used. The time evolution of the drop surface for different meshes is shown in Fig.~\ref{fig:grid_refinement}(a), where it can be observed that the results for different meshes are almost indistinguishable until $t^*=1.08$. At later times, the results for $L=12$ and 13 remain very close. The distribution of vaporization mass flux $j_\gamma^*$ on the drop surface at $t^*=0.72$ and 1.08 is shown in Figs.~\ref{fig:grid_refinement}(b) and (c). Here, $j_\gamma^*$ is plotted as a function of $\cos\theta$, where $\theta$ is the colatitude on the drop surface with respect to the origin, which is in the middle between the windward and leeward poles (see $t^*=0.72$ in Fig.~\ref{fig:grid_refinement}(a)). Again, the results for $L=12$ and 13 agree very well, confirming that the mesh resolution ($L=13$ or $D_0/\Delta=512$) used in the present study is sufficient to resolve the interfacial dynamics when vaporization is present.

%===============================================================================
%   Acknowledgments and bibliography
%=============================================================================== 
\section*{Acknowledgments}
This research was supported by ACS-PRF (\#62481-ND9) and NSF (\#1942324). We also acknowledge the ACCESS program for providing the computational resources that have contributed to the research results reported in this paper. The code used for the present simulations is an extension of the open-source multiphase flow solver \emph{Basilisk}, which is made available by St\'ephane Popinet and other collaborators.

\section*{Declaration of Interests}
The authors report no conflict of interest.

%\section*{References}
%\bibliographystyle{elsarticle-num}

%\bibliography{references}
%\bibliographystyle{jfm}

\end{document}